\documentclass[conference]{IEEEtran}

\usepackage{amsmath,amsfonts,bm}
\usepackage{algorithmic}
\usepackage{array}
\usepackage[caption=false,font=normalsize,labelfont=sf,textfont=sf]{subfig}
\usepackage{textcomp}
\usepackage{stfloats}
\usepackage{url}
\usepackage{verbatim}
\usepackage{graphicx}
\usepackage{textcomp}
\usepackage[dvipsnames]{xcolor}
\usepackage[linesnumbered,ruled,vlined]{algorithm2e}
\usepackage{hyperref}
\usepackage{tabulary}

\usepackage{float}

\usepackage{graphics}
\graphicspath{ {./figs/} }

\usepackage{caption}
\usepackage{subcaption}

\usepackage{wrapfig}

\usepackage[nobiblatex]{xurl}

\usepackage{amsthm}
\newtheoremstyle{mystyle} % name
  {3pt}   % Space above
  {3pt}
  {} % Body font
  {}      % Indent amount
  {} % Theorem head font
  {}     % Punctuation after theorem head
  { }     % Space after theorem head
  {}      % Theorem head spec

\theoremstyle{mystyle}

\usepackage{booktabs}

\usepackage{makecell}
\usepackage{multirow}

\usepackage{paralist}

\usepackage{booktabs}
\usepackage{siunitx}

\usepackage{mathtools,tikz,caption}
\DeclareRobustCommand\sampleline[1]{%
    \tikz\draw[#1] (0,0) (0,\the\dimexpr\fontdimen22\textfont2\relax)
    -- (2em,\the\dimexpr\fontdimen22\textfont2\relax);%
}

\usepackage{pifont}
\usepackage[most]{tcolorbox}
\tcbset{
  colback=green!5!white,
  colframe=green!85!black,
  boxsep=4pt,                 % No internal padding
  left=0pt, right=0pt,        % Remove left/right padding
  top=0pt, bottom=0pt,        % Remove top/bottom padding
  enhanced,
  before skip=2pt,
  after skip=3pt,
  sharp corners               % Optional: remove rounded corners
}

\usepackage{soul}
\newcounter{rqcounter}
\newcounter{subrqcounter}[rqcounter]
\definecolor{color-up-to-date}{HTML}{8CC5E3}
\definecolor{color-tood}{HTML}{2066A8}
\definecolor{color-pfet-tood}{HTML}{F72B8F}

\usepackage{lipsum}                     % Dummytext
\usepackage{xargs}                      % Use more than one optional parameter in a new commands
\usepackage[inline]{enumitem}
\setlist[enumerate]{nosep}

\usepackage[colorinlistoftodos,prependcaption,textsize=scriptsize]{todonotes}
\newcommandx{\unsure}[2][1=]{\todo[linecolor=red,backgroundcolor=red!25,bordercolor=red,#1]{#2}}
\newcommandx{\change}[2][1=]{\todo[linecolor=blue,backgroundcolor=blue!25,bordercolor=blue,#1]{#2}}
\newcommandx{\info}[2][1=]{\todo[linecolor=OliveGreen,backgroundcolor=OliveGreen!25,bordercolor=OliveGreen,#1]{#2}}
\newcommandx{\improvement}[2][1=]{\todo[linecolor=Plum,backgroundcolor=Plum!25,bordercolor=Plum,#1]{#2}}
\newcommandx{\thiswillnotshow}[2][1=]{\todo[disable,#1]{#2}}

\SetCommentSty{mycommfont}

\newcommand{\pkgdep}{{$<$package, dependency$>$} }

\newcommand{\mttu}{MTTU\textsubscript{dep}\xspace}
\newcommand{\mttr}{MTTR\textsubscript{dep}\xspace}
\newcommand{\MTTU}{Mean-Time-To-Update\textsubscript{dep}\xspace}
\newcommand{\MTTR}{Mean-Time-To-Remediate\textsubscript{dep}\xspace}

\newcommand\pkgname[1]{\textsf{\small #1}}
\newcommand\version[1]{$\operatorname{#1}$}

\newcommand{\npm}{\textsf{npm}\xspace}
\newcommand{\pypi}{\textsf{PyPI}\xspace}
\newcommand{\cratesio}{\textsf{crates.io}\xspace}

\newcommand{\datasetname}{\textsc{chrono\--resolution}\xspace}
\newcommand\dbtablename[1]{\textsc{#1}}
\newcommand{\highlight}[2]{%
    \vspace{.05\baselineskip}
    \colorlet{currentcolor}{.}%
    {\color{#1}%biolinum
    \noindent\fbox{\parbox{0.985\linewidth}{\color{currentcolor}#2}}}%
    \vspace{.05\baselineskip}
}

\newcommand{\goalstatement}{\textit{The goal of this paper is to aid practitioners and researchers in analyzing the state of the ecosystem dependency graph at release points using an enriched dataset with dependency resolution at release points for \npm, \pypi, and \cratesio packages.}\xspace}

\begin{document}

\title{\datasetname: A Dependency Resolution Dataset at Release Points for \npm, \pypi, and \cratesio Packages}

\author{
  \IEEEauthorblockN{Imranur Rahman, Jill Marley, Ranindya Paramitha, Laurie Williams}
  \IEEEauthorblockA{North Carolina State University\\
    Raleigh, NC, USA\\
    \{irahman3, jahmad5, rparami, lawilli3\}@ncsu.edu}
}

\maketitle

\begin{abstract}
% area/background
Dependency resolution at a specified point in time in the past can provide insight into software evolution in software ecosystems and facilitate the design of dynamic metrics (e.g., dependency freshness, dependency update rhythm).
% problem
However, dependency resolution at specified points in time is not possible in major software ecosystems due to a lack of support from package management tools.
% solution/goal statement
\goalstatement
% method
In this paper, we present a methodology to construct dependency resolution at release points of packages in software ecosystems, which we enrich with vulnerability data from OSV.
We apply our methodology to construct \datasetname, a dataset of dependency resolution at release points for \npm, \pypi, and \cratesio packages, and release it for future research.
\end{abstract}


\section{Introduction}
% area/background
Reusing software as dependencies improves developers' productivity~\cite{abdalkareem_why_2017}.
To facilitate software reuse, package registries, e.g., \npm for JavaScript, \pypi for Python, and \cratesio for Rust, store, manage, and distribute software packages and their dependencies.
However, the proliferation of packages in software registries makes software supply chain attacks via vulnerable dependencies an increasingly attractive attack vector~\cite{williams2025researchdirections}.
For example, \npm had 1M packages in 2019 and now has 3.67M packages as of 2025~\cite{depsdev}.

% problem
% The need for having historical dependency resolution
To better understand \pkgdep relationships and defend against software supply chain attacks through vulnerable dependencies, several researchers analyzed package version release information, \pkgdep relationships, and proposed metrics to measure package characteristics~\cite{liu_demystifying_2022,rahman2025whatspackagegettingvisibility,cox_measuring_2015,rahman2026quicklydevelopmentteamsupdate}.
For example, Rahman et al.~\cite{rahman2026quicklydevelopmentteamsupdate} designed two metrics, \MTTU (\mttu), and \MTTR (\mttr), to measure how quickly development teams update their vulnerable dependencies.
However, several of such metrics are dynamic, e.g., Common Vulnerabilities and Exposures (CVEs), dependency freshness, \mttu, and \mttr.
In addition, packages use dependency version constraints to specify which versions of dependencies they depend on.
For example, a package can specify \texttt{>=1.0.0} or \texttt{==1.0.0} as a version constraint for a dependency.
As a result, the version of the dependency that ends up installed in the end applications may change with different upstream releases.

% Limitations in the package management tools in doing historical resolution
To analyze the state of the package at release points, we need to resolve the dependency graph at some point in the past.
Release points are either the \verb|major|, \verb|minor|, or \verb|patch| releases made by a package.
Such dependency resolution at release points is non-trivial since we need to reconstruct the state of ecosystems (e.g., \npm, \pypi, and \cratesio) at those times.
\npm command-line tool has a \textit{time-traveling} feature (an undocumented \texttt{--before} argument found by Pinckney et al.~\cite{pinckney_large_2023}) to conduct a dependency resolution at specified points in time.
However, the \textit{time-traveling} feature does not exist for other major ecosystems (e.g., \pypi or \cratesio).
Research with dynamic metrics or involving longitudinal analysis would benefit from a dataset of dependency resolution at specified points in time for software ecosystems.

% solution/goal statement
\goalstatement
We release the dataset, construction scripts, and usage examples for future research.
Our dataset contains $146,651$ \npm, $48,608$ \pypi, and $15,690$ \cratesio packages with $2,221,947$ \npm, $360,969$ \pypi, and $158,849$ \cratesio\xspace \pkgdep relationships and $6,948$ security advisories.
% takeaway/how this is useful for others
Our provided dataset presents static dependency resolution from the past, and our provided methodology would allow future research to conduct dependency resolution at release points of their choice.

\begin{figure}
    \centering
    \includegraphics[width=.95\linewidth]{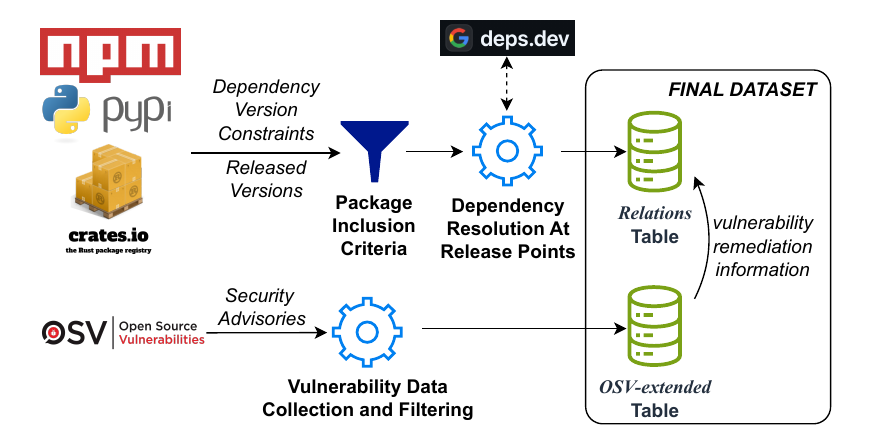}
    \caption{\datasetname construction workflow.}
    \label{fig:workflow}
\end{figure}

\begin{table*}[ht]
\centering
\small
\caption{Description of the \dbtablename{relations} table columns of \datasetname and how it was built.}
\resizebox{.9\linewidth}{!}{
\begin{tabular}{l | l | l} 
\toprule
Column name & Type & Description \\
\hline\hline
\multicolumn{3}{l}{\textbf{Initiated with data collection from \npm, \pypi, and \cratesio} (Section~\ref{subsec:collection}, row filtering in Section~\ref{subsec:inclusion})}\\\hline
ecosystem & String & The name of the ecosystem registry. \\
package & String & The name of the package in the ecosystem. \\
package version & SemVer & The version of the package in the ecosystem. \\
dependency & String & The dependency of the package. \\
dependency constraint & String & The version constraint set by the developers for the `dependency.' \\
constraint type & String & The type of the version constraint `dependency constraint' according to Rahman et al.~\cite{Rahman2025ASE}. \\ \hline
\multicolumn{3}{l}{\textbf{Added after dependency resolution at release points} (Section~\ref{subsec:depresol})}\\\hline
dependency version & SemVer & The resolved dependency version with the `dependency constraint' at the beginning of `Interval start.' \\
dependency highest version & SemVer & The highest available version of the dependency at the beginning of column `Interval start.' \\
Interval start & Timestamp & The start time of the interval for this \pkgdep relationship. \\
Interval end & Timestamp & The end time of the interval for this \pkgdep relationship. \\ \hline
\multicolumn{3}{l}{\textbf{Added in the data processing step, for \textit{remediated} using vulnerability data (Table~\ref{table:advisory-data})} (Section~\ref{subsec:dataproc})}\\\hline
updated & Boolean & If the `dependency version' matches with `dependency highest version' or not. \\
remediated & Boolean & If `dependency version' falls inside any CVE vulnerability ranges. \\
\bottomrule
\end{tabular}
}
\label{table:relations-data}
\end{table*}

\section{Related Work And The Originality Of The Dataset}
A variety of tools and services are available that store package metadata from software registries.
\texttt{Libraries.io}~\cite{libraries-io} hosts a detailed dataset of metadata across multiple package registries.
Similar websites are available, e.g., World of Code~\cite{ma2021world} and ecosyste.ms~\cite{ecosyste-ms}, to gather package metadata, source code, and dependency resolution from software registries.
However, no tools or websites exist for dependency resolution at release points in software registries. 
% Deps.dev~\cite{depsdev} provides information on the current dependency resolution, not the resolution of a dependency at a specific time.

% Papers involed in historical dependency resolution
Research has shown the importance of dependency resolution at specific points in time for software registries~\cite{jaime_goblin_2024,pinckney_large_2023,he_pinning_2025}.
Jaime et al.~\cite{jaime_goblin_2024} presented \texttt{GOBLIN} to ``rewind'' the dependency graph at some point in time 
% for Java Maven Central ecosystem packages 
to facilitate research on dynamic metrics (e.g., freshness~\cite{cox_measuring_2015}, rhythm~\cite{jaime_preliminary_2022}).
However, their dataset contains only Maven Central packages and is not suitable for cross-ecosystem analysis.
Other research used the time-traveling feature of \npm (with the \texttt{--before} argument) to conduct a large-scale analysis~\cite{pinckney_large_2023} and a simulation study~\cite{he_pinning_2025}.
% Pinckney et al.~\cite{pinckney_large_2023} built a time-traveling dependency resolver for \npm and conducted a large-scale analysis on the use of SemVer in \npm at specified points in time.
% They used the \npm command-line tool (with the \texttt{--before} argument) to conduct the time-traveling dependency resolution.
% He et al.~\cite{he_pinning_2025} used the time-traveling feature of \npm to conduct a simulation study of the cost and benefit of using pinning and floating.
However, no research or dataset is available for dependency resolution at release points for other major software ecosystems (e.g., the oldest ecosystem \pypi or one of the newest ecosystems \cratesio are often ignored).

% a description of the originality of the dataset (that is, even if the dataset has been used in a published paper, its complete description must be unpublished) and similar existing datasets (if any),
% \section{Originality Of The Dataset}
We performed two studies using the dataset presented in this paper.
For the first study, we designed two dependency update metrics, \MTTU (\mttu) and \MTTR (\mttr), to measure the responsiveness of development teams in keeping their dependencies up to date and mitigated~\cite{rahman2026quicklydevelopmentteamsupdate}.
Our designed metrics rely on dependency resolution data at release points to handle \emph{floating} version constraints correctly (e.g., automatically incorporating upstream releases) in our measurement.
The state of dependency graphs at release points of packages, e.g., which dependency version would have been installed using \emph{floating}, is a requirement of our designed metrics.
We then continue to our second study, in which we performed survival analysis to analyze the historical data split into time intervals to understand the impact of dependency version constraints (e.g., \emph{pinning} and \emph{floating}) on time to become outdated and time to become vulnerable dependencies~\cite{Rahman2025ASE}.

\textbf{What is new in this release.}
The internal snapshots used in~\cite{rahman2026quicklydevelopmentteamsupdate,Rahman2025ASE} were not publicly released and covered a narrower package set collected earlier.
This paper provides the first complete, publicly available description and release of the dataset: it extends coverage to all three ecosystems, adds the \emph{fix\_available} annotation for unfixed vulnerabilities, provides both raw and filtered variants, and releases the full data collection and transformation pipeline for reproducibility.
Researchers wishing to replicate or extend prior studies~\cite{rahman2026quicklydevelopmentteamsupdate,Rahman2025ASE} should use the \datasetname release.
% However, the completed description of the data is unpublished and we are filling the gap using this study by providing the complete description and sharing the historical dependency resolution dataset for \npm, \pypi, and \cratesio packages.

% if the data has been used by the authors or others, a description of how this was done including references to previously published papers,

% a description of the storage mechanism, including a schema if applicable,
\section{Dataset Construction Methodology}
In this section, we present our methodology for constructing the dataset of dependency resolution at release points. We illustrate the high-level workflow for the dataset construction process in Figure~\ref{fig:workflow}. Our dataset contains two tables: (1) \dbtablename{relations}, which contains the packages in the three ecosystems and their \pkgdep relationships at release points, and (2) \dbtablename{osv-extended}, which contains vulnerabilities gathered from OSV website~\cite{osv-dev} in the three ecosystems. We then used the vulnerability data from \dbtablename{osv-extended} to add the \textit{remediated} column in \dbtablename{relations}.

\subsection{\dbtablename{Relations}: Packages and Dependencies at Release Points}
In this section, we discuss how we constructed the \dbtablename{relations} table. The columns (and the step in which they are added) are shown in Table~\ref{table:relations-data}.  

\subsubsection{Package Metadata Collection}
\label{subsec:collection}
We collected \npm, \pypi, and \cratesio package metadata (version release information and \pkgdep relationships with version constraints for each version) on August 20, 2024.
Our required metadata are present in \texttt{package.json} file for \npm, \texttt{setup.py} or \texttt{pyproject.toml} for \pypi, and \texttt{Cargo.toml} for \cratesio packages.
% \myworries{one line about the PEP versioning and different package managers in PyPI.}
The JSON formatted package metadata are available at \href{https://registry.npmjs.org/express}{https://registry.npmjs.org/ $<$package-name$>$}, \href{https://pypi.org/pypi/requests/json}{https://pypi.org/pypi/$<$package-name$>$/json}, and \href{https://crates.io/api/v1/crates/serde}{https://crates.io/api/v1/crates/$<$crate-name$>$}.
We selected these three ecosystems for their diversity: \npm is the largest (with 3.67 million packages compared to Maven's 715k), \pypi is the oldest (introduced in 2003, whereas Maven Central came in 2005), and Cargo is the newest among the major software ecosystems.
We used deps.dev~\cite{depsdev} to collect the data for this phase, as also used in similar prior studies to collect package metadata in software ecosystems~\cite{hu_empirical_2024,akhoundali_morefixes_2024}.
% ~\cite{shen_understanding_2024,hu_empirical_2024,liu_detecting_2025,alhanahnah_depsrag_2024,akhoundali_morefixes_2024}.
At the end of this phase, we have $2,603,314$ \npm, $274,720$ \pypi, and $122,069$ \cratesio packages.
To validate data quality, the first author manually inspected a sample of 50 packages per ecosystem, comparing reported versions and \pkgdep relationships field-by-field against the live registry APIs and found 100\% match.

\subsubsection{Applying Package Inclusion Criteria}
\label{subsec:inclusion}
Before applying the inclusion criteria, we have 
% $3,000,103$ (
$2,603,314$ \npm, $274,720$ \pypi, and $122,069$ \cratesio packages.
Our package inclusion criteria are:
\begin{enumerate*}[label=(\roman*)]
    \item The package must be at least two years old, operationalized by checking the time difference between the first and latest available version release of the package. 
    \item The package must have at least one version release in the last two years.
    \item The package must have at least one dependency.
\end{enumerate*}

Our selection criteria are based on the work of Miller et al.~\cite{miller_understanding_2025}, who defined abandoned packages as those with two years of regular maintenance followed by two years of inactivity; we use this definition to exclude abandoned packages.
Additionally, the criterion of ``two years" is commonly used to determine whether a package is still actively maintained or not~\cite{li_comparison_2023,rahman2026quicklydevelopmentteamsupdate}. 
After applying our inclusion criteria, our resulting dataset contains $163,207$ ($146,651$ \npm, $48,608$ \pypi, and $15,690$ \cratesio) packages.
The filtering removed $2,456,663$ \npm ($\approx$94\%), $226,112$ \pypi ($\approx$82\%), and $106,379$ \cratesio ($\approx$87\%) packages.
We provide both the raw (unfiltered) and filtered versions of the dataset in the Zenodo repository~\cite{zenodo-artifact}, so that researchers with different inclusion criteria can work from the complete collection.
Packages excluded by the two-year age cutoff are of interest to researchers studying newer or feature-complete packages~\cite{coelho2017modern}.

\begin{table*}[ht]
    \centering
    \caption{An example of our dataset: \pkgname{hexo} package with one of its dependency \pkgname{moment} collected from Rahman et al.~\cite{Rahman2025ASE}.
    }
    \resizebox{\linewidth}{!}{
    \begin{tabular}{|c|c|c|c|c|c|c|c|c|c|c|c|c|}
    \hline
    \makecell[t]{row} & \makecell[t]{package} & \makecell[t]{package\\version} & \makecell[t]{dependency} & \makecell[t]{dependency\\constraint} & \makecell[t]{constraint\\type} & \makecell[t]{dependency\\version} & \makecell[t]{dependency\\highest version} & \textbf{\makecell[t]{Interval start}} & \textbf{\makecell[t]{Interval end}} 
    & \textit{\textbf{\makecell[t]{updated}}} & \textit{\textbf{\makecell[t]{remediated}}} 
    \\
    \hline\hline
    % 1 & {hexo} & $0.0.1$ & moment & $*$ & floating-major & $1.7.2$ & $1.7.2$ & $\operatorname{2012-10-07}$ & $\operatorname{2012-10-09}$ & true & true \\ \hline
     %& {$\ldots$} & $\ldots$ & $\ldots$ & $\ldots$ & $\ldots$ & $\ldots$ & $\ldots$ & $\ldots$ & $\ldots$ & $\ldots$ & $\ldots$  \\ \hline
    %116 & {hexo} & $3.0.1$ & moment & $2.9.0$ & pinning & $2.9.0$ & $2.9.0$ & $\operatorname{2015-04-06}$ & $\operatorname{2015-04-09}$ & true & true \\ \hline
    %117 & {hexo} & $3.0.1$ & moment & $2.9.0$ & pinning & $2.9.0$ & $2.10.2$ & $\operatorname{2015-04-09}$ & $\operatorname{2015-05-13}$ & \textcolor{red}{false} & true \\ \hline
    %118 & {hexo} & $3.0.1$ & moment & $2.9.0$ & pinning & $2.9.0$ & $2.10.3$ & $\operatorname{2015-05-13}$ & $\operatorname{2015-05-20}$ & \textcolor{red}{false} & true \\ \hline
    %119 & {hexo} & $3.1.0$ & moment & $\sim2.10.3$ & floating-patch & $2.10.3$ & $2.10.3$ & $\operatorname{2015-05-20}$ & $\operatorname{2015-05-20}$ & true & true \\ \hline
    % 120 & {hexo} & $3.1.1$ & moment & $\sim2.10.3$ & floating-patch & $2.10.3$ & $2.10.3$ & $\operatorname{2015-05-20}$ & $\operatorname{2015-07-26}$ & true & true \\ \hline
    % 121 & {hexo} & $3.1.1$ & moment & $\sim2.10.3$ & floating-patch & $2.10.5$ & $2.10.5$ & $\operatorname{2015-07-26}$ & $\operatorname{2015-07-28}$ & true & true \\ \hline
     & {$\ldots$} & $\ldots$ & $\ldots$ & $\ldots$ & $\ldots$ & $\ldots$ & $\ldots$ & $\ldots$ & $\ldots$ & $\ldots$ & $\ldots$  \\ \hline
    122 & {hexo} & $3.1.1$ & moment & $\sim2.10.3$ & floating-patch & $2.10.6$ & $2.10.6$ & $\operatorname{2015-07-28}$ & $\operatorname{2016-01-02}$ & true & true \\ \hline
    123 & {hexo} & $3.1.1$ & moment & $\sim2.10.3$ & floating-patch & $2.10.6$ & $2.11.0$ & $\operatorname{2016-01-02}$ & $\operatorname{2016-01-09}$ & \textcolor{red}{false} & true \\ \hline
    124 & {hexo} & $3.1.1$ & moment & $\sim2.10.3$ & floating-patch & $2.10.6$ & $2.11.1$ & $\operatorname{2016-01-09}$ & $\operatorname{2016-02-03}$ & \textcolor{red}{false} & true \\ \hline
    125 & {hexo} & $3.1.1$ & moment & $\sim2.10.3$ & floating-patch & $2.10.6$ & $2.11.2$ & $\operatorname{2016-02-03}$ & $\operatorname{2016-02-28}$ & \textcolor{red}{false} & \textcolor{red}{false} \\ \hline
    126 & {hexo} & $3.2.0$ & moment & $\sim2.11.2$ & floating-patch & $2.11.2$ & $2.11.2$ & $\operatorname{2016-02-28}$ & $\operatorname{2016-03-07}$ & true & true \\ \hline
    % the last two columns are negated from what is in the dataset since the column names are negated.
    
    {$\ldots$} & $\ldots$ & $\ldots$ & $\ldots$ & $\ldots$ & $\ldots$ & $\ldots$ & $\ldots$ & $\ldots$ & $\ldots$ & $\ldots$ & $\ldots$ \\
    % \hline
    \end {tabular}
    }
    \label{table:running-example}
\end{table*}

\subsubsection{Dependency Resolution At Release Points}
\label{subsec:depresol}
After collecting the data and applying the inclusion criteria, we divide each \pkgdep relationship into multiple time intervals, based on release points.
Release points refer to the points in time when a \texttt{major}, \texttt{minor}, or \texttt{patch} release was made by the package or the dependency.
Time intervals refer to the time between two release points for a $<$package, dependency$>$.
According to our definition of time intervals, no new versions of either the package or its dependencies are released during each interval.
Table~\ref{table:running-example} shows an example of time intervals: `interval start' and `interval end'. %An example of these intervals, labeled as `interval start' and `interval end', is shown in Table~\ref{table:running-example}.
% In each interval, the dependency version constraint is resolved based on the versions of dependencies available at the beginning of the interval to ensure that the resolution reflects the historical state.
%We use deps.dev~\cite{depsdev} to perform the dependency resolution under this constraint and populate the `dependency version' column in Table~\ref{table:running-example}.

As noted by previous research~\cite{pinckney_large_2023,pinckney_npm-follower_2023,he_pinning_2025}, \npm offers a ``time-travel'' feature (the \texttt{--before} argument), which allows resolving dependencies at specified points in time.
However, \pypi and \cratesio do not provide this feature.
Thus, we contacted the Google deps.dev team, who provided us with the dependency resolution at release times for the packages in our dataset (snapshot date: August 20, 2024).
\textit{Note that the dataset cannot be rebuilt via public deps.dev API, and it will require special access from Google.}
To verify deps.dev's resolution validity, the first author manually performed dependency resolution for the same 50 packages per ecosystem and found that deps.dev matched exactly (see Section~\ref{sec:threats} for caveats on PyPI pre-2020 data).
We exclude SemVer pre-release and build-metadata qualifiers from our dependency resolution, as package managers ignore them unless explicitly pinned.

\subsection{\dbtablename{OSV-extended}: Vulnerability Data}
We gathered security advisories (CVE data) from the Google Open Source Vulnerabilities (OSV) database, available at osv.dev~\cite{osv-dev}, for \npm, \pypi, and \cratesio packages on September 12, 2024 (OSV data cutoff).
We selected OSV because it aggregates data from various vulnerability feeds~\cite{osv-data-sources}, such as GitHub Security Advisories~\cite{github-advisory-database}, PyPA, and GoVulDB, across multiple ecosystems, and presents the data in a standardized OSV format.
After collecting all security advisories for the relevant ecosystems, we filtered out advisories for which the affected package is not in our dataset.
Unlike prior work, we \emph{retain} advisories for which no fixed version was available at the time of data collection and annotate them with a Boolean \emph{fix\_available} column (Table~\ref{table:advisory-data}).
Retaining unfixed vulnerabilities allows security researchers to study long-lived and unpatched vulnerabilities; the \emph{fix\_available} flag enables downstream analyses to distinguish fixable from unfixable exposures.
The resulting dataset after filtering for relevance to our package set consisted of $2,192$ \npm, $3,767$ \pypi, and $989$ \cratesio vulnerabilities with a fixed version, plus additional advisories without a known fix annotated as \emph{fix\_available} = \textcolor{red}{false}.
Next, we converted the data to an SQL table with \emph{vul\_id} (advisory identifier), \emph{ecosystem}, \emph{package} (vulnerable package name), \emph{vul\_introduced} (version where the vulnerability was introduced), \emph{vul\_fixed} (fixed version, or NULL if no fix exists), and \emph{fix\_available} (Boolean), as shown in Table~\ref{table:advisory-data}.
If a vulnerability contained multiple vulnerable version ranges, we separated it into multiple SQL rows, each corresponding to one SemVer vulnerable version range, to facilitate analysis.

Since vulnerability data is often inconsistent~\cite{croft2023dataquality}, we post-processed versions to ensure SemVer~\cite{semver-spec} compliance (\textit{major.minor.patch}): implied components were filled in (e.g., \version{0} $\to$ \version{0.0.0}), and rows with non-conforming versions (e.g., extra components, non-numeric identifiers) were removed. 

\subsection{Final Dataset Processing}

\subsubsection{Data Cleaning}
Deps.dev resolves the full dependency tree, creating duplicates.
We keep only direct dependencies to eliminate duplicates; the full transitive graph can be reconstructed by iteratively self-joining \dbtablename{relations} on overlapping intervals.
A reference reconstruction script is provided in the replication package.
Then, we used the \verb|semver|~\cite{postgres-semver} extension, an implementation of SemVer, to convert the data types of our tables that contain \texttt{versions} of packages or dependencies, removing any rows where the version could not be converted.
For example, \verb|semver| extension can handle 31-bit integer for each of the \texttt{major}, \texttt{minor}, or \texttt{patch} values.
Some packages' version (e.g., \version{1.0.20230603010803}) resulted in errors (e.g., bad SemVer value) because of overflowing the \texttt{patch} with \version{20230603010803}.

\begin{table*}[t]
\centering
\small
\caption{Description of the \dbtablename{osv-extended} table columns of \datasetname.}
% \resizebox{\columnwidth}{!}{
\begin{tabular}{l | l | l}
\hline
Column name & Type & Description \\
\hline\hline
vul\_id & String & The unique identifier for the CVE. \\
ecosystem & String & The ecosystem of the affected package for the CVE. \\
package & String & The name of the vulnerable package for the ecosystem for the CVE. \\
vul\_introduced & SemVer & Version of the `package\_name' where the vulnerability was introduced. \\
vul\_fixed & SemVer (nullable) & Fixed version of the `package\_name' for the CVE; NULL if no fix was available at data collection. \\
fix\_available & Boolean & Whether a fixed version existed at the OSV data collection date (September 12, 2024). \\
\hline
\end{tabular}
% }
\label{table:advisory-data}
\end{table*}

\begin{table}[t]
\centering
\small
\caption{Statistics of \datasetname.}
% \resizebox{\columnwidth}{!}{
% \begin{tabular}{||p{0.15\linewidth} | p{0.19\linewidth} | p{0.23\linewidth} | p{0.23\linewidth}||} 
\begin{tabular}{ l | r | r| r } 
\hline
\makecell[t]{Ecosystem} & \makecell[t]{Number\\ of unique \\packages} & \makecell[t]{Number of \\$<$package\\dependency$>$\\ relationships} & \makecell[t]{Number \\ of unique \\$<$package\\dependency$>$\\ relationships} \\
\hline\hline
\cratesio & $15,690$ & $8,070,357$ & $158,849$ \\
\npm & $146,651$ & $245,307,373$ & $2,221,947$ \\
\pypi & $48,608$ & $20,859,740$ & $360,969$ \\
\hline
\end{tabular}
% }
\label{table:dataset-stat}
\end{table}

\subsubsection{Data Processing}
\label{subsec:dataproc}
%After conducting the dependency resolution at release points and enriching with vulnerability data,
We next populated the Boolean columns `updated' and `remediated' in Table~\ref{table:running-example} to indicate whether the package uses an outdated or vulnerable dependency.
A resolved dependency version was marked as outdated (updated = \textcolor{red}{false}) if it is not the highest SemVer version of the dependency available at the start of the interval.
Similarly, a resolved dependency version was marked as vulnerable (remediated = \textcolor{red}{false}) if it falls within the range of vulnerable versions for a given vulnerability (Table~\ref{table:advisory-data}) \emph{and} a fixed version was available at the start of the interval (\emph{fix\_available} = true in Table~\ref{table:advisory-data}).
If a package used a vulnerable dependency version at a time interval but no fixed version had been released at that time, we consider that the package is not at fault and mark the row as remediated = true.
Researchers studying long-lived unpatched exposures can identify relevant rows by joining \dbtablename{relations} with \dbtablename{osv-extended} on \emph{fix\_available} = false.

% \subsubsection{Example Usage}
% \label{subsec:usage}
% To lower the barrier for new users, we illustrate two representative SQL queries.

% \noindent\textbf{(1) Outdated dependencies at a given date.}
% \begin{verbatim}
% SELECT package, dependency, dependency_version,
%        dependency_highest_version, interval_start
% FROM relations
% WHERE updated = false
%   AND interval_start <= '2023-01-01'
%   AND interval_end   >  '2023-01-01'
%   AND ecosystem = 'npm';
% \end{verbatim}

% \noindent\textbf{(2) Packages exposed to long-lived unfixed vulnerabilities.}
% \begin{verbatim}
% SELECT r.package, r.dependency,
%        o.vul_id, o.vul_introduced
% FROM relations r
% JOIN osv_extended o
%   ON r.ecosystem = o.ecosystem
%  AND r.dependency = o.package
%  AND r.dependency_version >= o.vul_introduced
%  AND o.fix_available = false
% WHERE r.interval_start <= '2023-01-01'
%   AND r.interval_end   >  '2023-01-01';
% \end{verbatim}

% Additional end-to-end usage examples and a Python notebook are provided in the Zenodo replication package~\cite{zenodo-artifact}.

\section{Applicability}
Researchers studying evolution of ecosystems through dependencies and analyzing dynamic metrics (e.g., dependency freshness, update rhythm, \MTTU, \MTTR) can use our dataset.
However, our dataset is unsuitable for research on new, abandoned, taken-down, short-lived, or typosquatted packages since our dataset represents older, active packages with at least one dependency.

\highlight{black}{\textbf{Final Dataset:}
Our final dataset, \datasetname, contains two schemas and data dump of \dbtablename{relations} (Table~\ref{table:relations-data}) and \dbtablename{osv-extended} (Table~\ref{table:advisory-data}).
The statistics of \datasetname are presented in Table~\ref{table:dataset-stat}.
We share the schema, data dump (filtered and raw), all data collection and transformation scripts, and example SQL queries in a \textbf{Zenodo repository~\cite{zenodo-artifact}}.
The replication package documents the exact API endpoints, query structures, and snapshot dates used for deps.dev and OSV, and includes instructions for re-running or adapting the pipeline as external APIs evolve.
}

% ideas for future research questions that could be answered using the dataset,
% ideas for further improvements that could be made to the dataset, and
% any limitations and/or challenges in creating or using the dataset.
\section{Future Research Ideas}
The research directions below require release-point dependency resolution specifically\textemdash{}not just ecosystem-level metadata\textemdash{}which is the unique feature of \datasetname.

\begin{description}[leftmargin=1em]
    \item[Ecosystem evolution.] Longitudinal analysis of how the active dependency graph changes version by version\textemdash{}tracking which packages entered or left common use over time\textemdash{}requires release-point resolution that static snapshots cannot provide.
    \item[Software measurement.] Clustering packages by dependency graph structure and tracking developers' centrality trends across ecosystems over time would give researchers a better understanding of the project and the developers' centrality trends across ecosystems.
    % Our provided dataset enables computing historical centrality metrics without re-running resolution.
    \item[Version constraint and dependency update.] How the failure of certain projects (e.g., a log4j-style event) cascades through the ecosystem, whether projects systematically avoid major updates, and what version constraints projects prefer for specific dependencies are three potential research questions for future researchers.
    \item[Vulnerability.] The proportion of vulnerabilities present in the dependency network at a specified time, risk of n-day exposure, and study of long-lived unpatched vulnerabilities using the \emph{fix\_available} = false rows in \dbtablename{osv-extended} are three potential research directions.
\end{description}
% the other ideas

% transitive

% For practitioners, tool builders, researchers

\section{Threats to Validity and Limitations}
\label{sec:threats}

\noindent\textbf{\emph{Construct Validity: Retroactive SemVer Resolution for PyPI.}}
Deps.dev applies a modern, backtracking SemVer resolver retroactively; for \pypi packages released before pip 20.3 (November 2020), resolved versions represent what the \emph{modern} resolver would have installed, which may differ from what developers actually experienced.
% Researchers should treat pre-2020 \pypi results as mathematically correct under modern SemVer semantics, not ground-truth historical installations.
So we recommend limiting \pypi analyses to post-2020 data or interpreting earlier data only as reconstructions.
Future work may quantify this discrepancy using period-accurate pip binaries.

\noindent\textbf{\emph{Construct Validity: Validation Sample Size.}}
Our manual validation covered 50 packages per ecosystem via field-by-field API comparisons, not end-to-end reproduction with historical package manager binaries.
Our validation is a smoke test of the pipeline and dependency resolution rather than statistical evidence of the correctness across all 163k packages.
Future efforts should expand to a statistically defensible sample and reproduce resolutions using containerized, period-accurate package manager versions.
In addition, our validation used live registry API and modern resolver behavior from deps.dev, so it cannot detect retroactive resolution errors.
A stratified sampling based validation with period-accurate tooling (e.g., npm \texttt{--before} and use of historical pip binaries) is a promising direction for future work.

\noindent\textbf{\emph{Package Inclusion.}}
The two-year age cutoff~\cite{miller_understanding_2025} may exclude younger or feature-complete packages~\cite{coelho2017modern}.
This threshold is a design choice and can be overridden via our released pipeline.
% As a result, our dataset represents older, active packages with at least one dependency, rather than a general snapshot of the ecosystems.

\noindent\textbf{\emph{Vulnerability Data Coverage.}}
OSV data was collected on September 12, 2024; later disclosures are not reflected.
OSV aggregates multiple feeds~\cite{osv-data-sources}, but vulnerabilities absent from those feeds do not appear in our dataset.

% \section{Conclusion}
% We present \datasetname, a dataset of dependency resolution at release points for \npm, \pypi, and \cratesio packages, along with its construction methodology, replication scripts, and usage examples.

% \noindent\textbf{\emph{Data Availability}}
% We share the schema and the data dump of our dataset in a Zenodo repository~\cite{zenodo-artifact}.
% More information about the dataset storage are included in the repository.

\section*{Acknowledgment}
This work was supported and funded by the National Science Foundation Grant No.\ 2207008 and Google.
Any opinions expressed in this material are those of the author(s) and do not necessarily reflect the views of the National Science Foundation or Google.

% \bibliographystyle{abbrv}
\bibliographystyle{acm}
\bibliography{websites,references}

\end{document}